\documentstyle[aps,amsmath,multicol,prl,epsfig]{revtex}

\begin{document}
\title{Resonant and Non-Resonant Modulated Amplitude Waves for Binary
Bose-Einstein Condensates in Optical Lattices}
\author{Mason A. Porter$^1$, P.G. Kevrekidis$^2$, and Boris A. Malomed$^3$}
\address{$^1$School of Mathematics and Center for Nonlinear Science, \\
Georgia Institute of Technology, Atlanta GA 30332, USA \\
$^2$Department of Mathematics and Statistics,\\
University of Massachusetts, Amherst MA 01003-4515, USA \\
$^3$Department of Interdisciplinary Studies, Faculty of Engineering, \\
Tel Aviv University, Tel Aviv 69978, Israel}
\date{\today }
\maketitle

\begin{abstract}
We consider a system of two Gross-Pitaevskii (GP) equations, in the presence
of an optical-lattice (OL) potential, coupled by both nonlinear and linear
terms. This system describes a Bose-Einstein condensate (BEC) composed of
two different spin states of the same atomic species, which interact
linearly through a resonant electromagnetic field. In the absence of the OL,
we find plane-wave solutions and examine their stability. In the
presence of the OL, we derive a system of amplitude equations for spatially
modulated states which are coupled to the periodic potential through the
lowest-order subharmonic resonance. We determine this averaged system's
equilibria, which represent spatially periodic solutions, and subsequently
examine the stability of the corresponding solutions with direct simulations
of the coupled GP equations. We find that symmetric (equal-amplitude) and
asymmetric (unequal-amplitude) dual-mode resonant states are, respectively,
stable and unstable. The unstable states generate periodic oscillations
between the two condensate components, which is possible only because of the
linear coupling between them.  We also find four-mode states, but they are
always unstable.  Finally, we briefly consider ternary (three-component) 
condensates.
\end{abstract}

\vspace{2mm}


\section{Introduction}

At sufficiently low temperatures, particles in a dilute boson gas condense in
the ground state, forming a Bose-Einstein condensate (BEC).
This was first observed experimentally in 1995 in Na and Rb
vapors\cite {pethick,stringari,ketter,edwards}.

In the mean-field approximation, a dilute BEC is described by the
nonlinear Schr\"{o}dinger (NLS) equation with an external
potential, which is also called the Gross-Pitaevskii (GP)
equation. In particular, BECs may be considered in the
quasi-one-dimensional (quasi-1D) regime, with the transverse
dimensions of the condensate on the order of its mean healing
length $\chi $ (given by $\chi ^{2}=(8\pi n|a|)^{-1}$) and a much larger
longitudinal dimension \cite{bronski,bronskirep,bronskiatt,stringari}.
The length $\chi $ is determined by the mean atomic density $n$ and the
two-body $s$-wave scattering length $a$, where the interactions between
atoms are repulsive if $a>0$ and attractive if
$a<0$ \cite{pethick,stringari,kohler,baiz}.

The quasi-1D regime, which corresponds to \textquotedblleft
cigar-shaped\textquotedblright\ BECs, is described by the 1D limit of the 3D
mean-field theory (rather than by a 1D mean-field theory proper, which would
only be appropriate for extremely small transverse dimensions of order
$\sim a$) \cite{bronski,bronskiatt,bronskirep,salasnich,towers}. In this
situation, the condensate wave function $\psi (x,t)$ obeys the effective 1D GP
 equation,
\[
i\hbar \psi _{t}=-\frac{\hbar ^{2}}{2m}\psi _{xx}+g|\psi |^{2}\psi
+V(x)\psi ,
\]
where $m$ is the atomic mass, $V(x)$ is an external potential,
$g=[4\pi \hbar ^{2}a/m][1+O(\eta ^{2})]$, and $\eta
=\sqrt{n|a|^{3}}$ is the dilute-gas
parameter\cite{stringari,kohler,baiz,band}. Experimentally
relevant potentials $V(x)$ include harmonic traps and periodic
potentials (created as optical lattices, which are denoted OLs and arise as
interference
patterns produced by coherent counterpropagating
laser beams illuminating the condensate).  In the presence of both potentials, 
$ V(x)=V_{0}\cos \left[
2\kappa (x-x_{0})\right] +V_{1}x^{2}/2$, where $x_{0}$ is the
offset of the periodic potential relative to the center of the the
parabolic trap. When $\left( 2\pi /\kappa \right) ^{2}V_{1}\ll
V_{0}$, the potential is dominated by its periodic component over
many periods \cite {kutz,promislow,lattice}; for example, when
$V_{0}/V_{1}=500$ and $\kappa =10 $, the parabolic component in
$V(x)$ is negligible for the 10 periods closest to the trap's center.
In this work, we set $V_{1}=0$ and focus entirely on OL
potentials. This assumption is motivated by numerous recent
experimental studies of BECs in OLs \cite{hagley,anderson} and is widely
adopted in theoretical studies \cite
{bronski,bronskiatt,bronskirep,space1,space2,promislow,kutz,malopt,alf,smer,mapbecprl,mapbec,mueller,wu4,machholm,pethick2}.

Multiple-component BECs, which constitute the subject of this
work, have been considered in {a number of} theoretical works
\cite {couple1,couple2,couple3,couple4,couple5,couple6,dec}.
Mixtures of two different species (such as $^{85}$Rb and
$^{87}$Rb) are described by nonlinearly coupled GP equations:
\begin{align}
i\hbar \frac{\partial \psi _{1}}{\partial t}& =-\frac{\hbar
^{2}}{2m_{1}} \nabla ^{2}\psi _{1}+g_{1}|\psi _{1}|^{2}\psi
_{1}+V(x)\psi _{1}+h|\psi
_{2}|^{2}\psi _{1},  \nonumber \\
i\hbar \frac{\partial \psi _{2}}{\partial t}& =-\frac{\hbar
^{2}}{2m_{2}} \nabla ^{2}\psi _{2}+g_{2}|\psi _{2}|^{2}\psi
_{2}+V(x)\psi _{2}+h|\psi _{1}|^{2}\psi _{2},  \label{cnls1}
\end{align}
where $m_{1,2}$ are the atomic masses of the species, $g_{j}\equiv 4\pi
\hbar ^{2}a_{j}/m_{j}$ corresponds to the self-scattering length $a_{j}$,
and
\begin{equation}
h\equiv g_{12}=2\pi \hbar ^{2}a_{12}\left( m_{1}+m_{2}\right) /\left(
m_{1}m_{2}\right)  \label{h}
\end{equation}
depends on the cross-scattering length $a_{12}$
\cite{couple2}. There are numerous subcases of Eqs. (\ref{cnls1})
to consider, as various combinations of signs for the scattering
coefficients $g_{1}$, $g_{2}$, and $ h$ may occur. It is important
to note, however, that if $g_{1,2}$ are positive (repulsion
between the atoms), then $h$ is normally positive as well. However,
if $g_{1,2}$ are negative (corresponding to the less typical
case of attraction
between atoms belonging to a single species), then $h$ [see Eq. (\ref{h})]
may be {\it either} positive or negative.

The system (\ref{cnls1}) resembles a well-known model
describing the nonlinear self-phase-modulation (SPM) and
cross-phase-modulation (XPM) interactions of light waves with
different polarizations or carried by different wavelengths
in nonlinear optics \cite{Agrawal}. In the case of optical fibers,
the evolution variable is the propagation distance $z$ (rather
than time), and the role of $x$ is played by the reduced temporal
variable $\tau $ \cite{Agrawal}. In optical models, however, the
choice of the nonlinear coefficients is limited to the
combinations $g_{1}=g_{2}=3h/2$ for orthogonal linear
polarizations in a birefringent fiber and $ g_{1}=g_{2}=h/2$ for
circular polarizations or different carrier wavelengths. In fact,
the latter case is quite important in the application to
two-component BECs as well, as it occurs if one assumes that the
collision lengths for interactions between all the atoms are the
same.

Another physically interesting feature, which we include in the model to be
considered below, is linear coupling between the two wave functions. This
occurs in a mixture of two different spin states of the
same isotope, which arises through a resonant microwave field that
induces transitions between the
states\cite{linear-coupling1,linear-coupling2}. Condensates
containing two different spin states of $^{87}$Rb have been created
experimentally via sympathetic cooling \cite{myatt}. In this
situation, the normalized coupled GP equations take the form
\begin{align}
i\frac{\partial \psi _{1}}{\partial t}& =-\nabla ^{2}\psi _{1}+g|\psi
_{1}|^{2}\psi _{1}+V(x)\psi _{1}+h|\psi _{2}|^{2}\psi _{1}+\alpha \psi _{2},
\nonumber \\
i\frac{\partial \psi _{2}}{\partial t}& =-\nabla ^{2}\psi _{2}+g|\psi
_{2}|^{2}\psi _{2}+V(x)\psi _{2}+h|\psi _{1}|^{2}\psi _{2}+\alpha \psi _{1},
\label{cnls2}
\end{align}
where the self- and cross-scattering coefficients are ${ g_1 =
g_2\equiv g}$ and $h$, and the linear coupling coefficient is
$\alpha $, which can always be made positive without loss of
generality.

Experimental studies of mixtures of two interconvertible
condensates (with positive scattering lengths) loaded in an OL
have not yet been reported. However, all the necessary
experimental ingredients for such a work are currently available.
Moreover, in a very recent paper \cite{kuklov}, an experimental
procedure, based on Ramsey spectroscopy and adjusted exactly
for such a system, was elaborated. Experiments in this setting
would be quite interesting, as they would allow the study of the
direct interplay between two crucially important physical
factors used as tools in the current experimental work---namely,
the OL potential and inter-conversion between two different spin
states in the BEC, controlled by the resonant field.  Furthermore, 
there are now recent experimental results with linearly
coupled BECs \cite{dsh}. 
The use of an optical potential in the latter setting is a rather
straightforward extension.

The model combining nonlinear XPM and linear couplings, as in Eqs.
(\ref {cnls2}), occurs in fiber optics as well. In that case, the
linear coupling is generated by a twist applied to the fiber in
the case of two linear polarizations, and by an elliptic
deformation of the fiber's core in the case of circular
polarizations (see, for example, \cite{old}). However,\ linear
coupling is impossible in the case of two different wavelengths.
Another optical model, with only linear coupling between two
modes, applies to dual-core nonlinear fibers, as discussed in
\cite{UNSW} (and references therein). In the context of BECs, it
may correspond to a special case in which the cross-scattering
length is made (very close to) zero using a Feshbach resonance \cite{inouye}.

In this work, we aim to investigate modulated-wave states in the binary BEC
described by Eqs. (\ref{cnls2}), which include both nonlinear and linear
couplings and an OL potential. We stress that the interplay between the
microwave-induced linear coupling in the binary model and the OL-induced
periodic potential has never before been considered. As both features
represent important laboratory tools, the results reported here suggest
possibilities for new experiments. The model we study predicts new dynamical 
effects, such as oscillations of matter between the linearly-coupled components
trapped in the potential wells of the OL.

In our study, we begin by examining plane-wave solutions with
$V(x)\equiv 0$. When $V(x)\neq 0$, we apply a
standing-wave ansatz to Eq. (\ref{cnls2}), which leads to a system of coupled
 parametrically forced Duffing equations describing the spatial evolution of
the fields.  Using the method of averaging \cite{675,rand}, we study periodic
 solutions of the latter system (called \textquotedblleft modulated amplitude
waves\textquotedblright and denoted MAWs).  The stability of MAWs
(and the ensuing dynamics, in the case of instability) is then
tested by numerically simulating the underlying system of coupled
GP equations. This approach, though simpler than the more
``rigorous''  computation of linear stability eigenvalues for
infinitesimal perturbations, provides a more realistic emulation of 
physical experiments. Note additionally that although our stability 
results will be illustrated by a few selected examples, we have 
checked---by exploring different parameter regions---that these 
examples represent the MAW stability features rather generally.

The MAW solutions are especially interesting when the system
exhibits a spatial resonance. In this work, we consider both
non-resonant solutions and solutions featuring a subharmonic
resonance of the $2\!:\!1\!:\!1$ form. The latter situation has
been studied in the context of period-doubling in {\em
single-component} BECs in an OL potential
\cite{pethick2,mapbecprl,mapbec}, but---to the best of our
knowledge---spatial-resonance states in models of composite BECs
have not been considered previously.

An alternative (but less general) approach to the study of binary BECs with
linear coupling, loaded into an OL, would be to seek exact
elliptic-function solutions to Eqs. (\ref{cnls2}) for the case of
elliptic-function potentials, $V(x)=-V_{0}{\rm sn}^{2}(\kappa x,k)$, as
has been done earlier in the two-component model without linear coupling
\cite{dec}. In that work, stable standing-wave solutions were found under the 
assumption that the interaction matrix is positive definite. This occurs, for
instance, when all the interactions are repulsive, although small negative
cross-interactions are compatible with this condition as well.

The rest of this paper is structured as follows: In section II, we derive
plane-wave solutions and analyze their stability. In section III, we introduce
modulated amplitude waves, and in section IV, we derive and solve averaged
equations that describe them in both non-resonant and resonant situations.
We corroborate our results and test the stability of the MAWs using
numerical simulations. In section V, we briefly examine a more general model
of a ternary (three-component) BEC with linear couplings. Finally, we
summarize our results in section VI.

\section{Plane-Wave Solutions}

In the absence of the external potential ($V=0$), we find plane-wave
solutions of the form
\begin{equation}
\psi _{j}=R_{j}\exp \left[ i(k_{j}x-\mu _{j}t)\right] \,,\quad j=1,2\,.
\label{plane}
\end{equation}
For the linearly coupled GP equation (\ref{cnls2}), it is
necessary that $ k_{1}=k_{2}\equiv k$ and $\mu _{1}=\mu
_{2}\equiv \mu $. Without linear coupling, [as in Eqs.
(\ref{cnls1})], one may seek a broader class of solutions with
independent frequencies (which correspond to chemical potentials in the
physical context of BECs).  It is important to note that the results obtained
 in the study of optical models suggest that the addition of linear
coupling terms to a system of coupled NLS equations drastically alters
the dynamical behavior \cite{old}. In this section, we study the
model without the OL, which will be included in subsequent sections.

Inserting Eq. (\ref{plane}) into Eqs. (\ref{cnls2}) yields
\begin{align}
\mu R_{1}& =k^{2}R_{1}+gR_{1}^{3}+hR_{1}R_{2}^{2}+\alpha R_{2}\,, \\
\mu R_{2}& =k^{2}R_{2}+gR_{2}^{3}+hR_{1}^{2}R_{2}+\alpha R_{1}\,,
\end{align}
which implies that
\[
\lbrack (g-h)R_{1}R_{2}-\alpha ][R_{1}^{2}-R_{2}^{2}]=0\,.
\]
One of the following two relations must then be satisfied:
\begin{equation}
R_{1}=\pm R_{2}\,;  \label{plane1}
\end{equation}
\begin{equation}
R_{1}R_{2}=\frac{\alpha }{g-h}\,.  \label{plane2}
\end{equation}

With Eq. (\ref{plane1}), a nonzero solution satisfies $R_{2}=\pm
\sqrt{\left( k^{2}+\mu \mp \alpha \right) /\left( g+h\right)
}$. It exists when $g+h>0$, provided $k^{2}+\mu \mp \alpha
>0$, and when $g+h<0$, if $ k^{2}+\mu \mp \alpha <0$. When
$g=-h$, one obtains solutions of the form $ \left(
R_{1},R_{2}\right) =\left( \pm R,R\right) $ with arbitrary $R$ and
$ k^{2}=\mu \mp \alpha $.

From Eq. (\ref{plane2}), one finds that
\begin{equation}
R_{2}^{2}=\frac{\mu -k^{2}}{2g}\pm \frac{1}{2}\sqrt{\left( \frac{\mu
-k^{2}}{g}\right) ^{2}-\frac{4\alpha ^{2}}{(g-h)^{2}}}\,,  \label{tip2}
\end{equation}
under the restriction that this expression must be positive. When
$\alpha \neq 0$, the term inside the square root is smaller in
magnitude than the one outside, so solutions of this type exist as
long as $\left( \mu -k^{2}\right) g\geq 0$ and the argument of
the square root in (\ref{tip2}) is non-negative. Hence, for
repulsive and attractive BECs, respectively, the first condition implies, 
$\mu >k^{2}$ and $\mu <k^{2}$. When $ h=0$,
Eq. (\ref{tip2}) takes the form
\[
R_{2}^{2}=\left( 2g\right) ^{-1}\left( \mu -k^{2}\pm \sqrt{(\mu
-k^{2})^{2}-4\alpha ^{2}}\right) \,.
\]
For both $h=0$ and $h=2g$, the condition on the argument of the square root
implies that to obtain real solutions, it is necessary to impose the
condition $|\mu -k^{2}|~\geq 2\alpha $.

To examine the stability of the plane waves, we substitute
\[
\psi _{j}(x,t)=\phi _{j}(x,t)[1+\epsilon _{j}(x,t)]\,,\quad |\epsilon
_{j}|^{2}\ll 1\,,
\]
into Eqs. (\ref{cnls2}). This yields coupled linearized equations
for $ \epsilon _{1}(x,t)$ and $\epsilon _{2}(x,t)$. Assuming that
$\epsilon _{j}$ is periodic in $x$, it can be expanded in a
Fourier series,
\[
\epsilon _{j}(x,t)=\sum_{n=-\infty }^{\infty }\hat{\epsilon}_{jn}(t)\exp
(i\nu _{n}x)\,,
\]
where the $n$th mode has wavenumber $\nu _{n}$. The perturbation growth
rates that determine the stability of the $n$th mode are then given by
\begin{equation}
\lambda _{n}=-2ik\nu _{n}\pm \nu _{n}\sqrt{-\nu
_{n}^{2}-g(|R_{1}|^{2}+|R_{2}|^{2})\pm \sqrt{
g^{2}(|R_{1}|^{2}-|R_{2}|^{2})^{2}+4h^{2}|R_{1}|^{2}|R_{2}|^{2}}}\,,
\label{lambda}
\end{equation}
where the two sign combinations $\pm $ are independent (so there
are four distinct eigenvalues). Instability occurs when the
expression under the square root in Eq. (\ref{lambda}) has a
positive real part, causing the side-band modes $k+\nu _{n}$,
$k-\nu _{n}$ of the perturbed solution to grow exponentially. In
single-component condensates, this can occur only for $ g<0$
\cite{split}.

Eigenvalues whose interior square root in Eq. (\ref{lambda}) has a $+$ sign
will produce the instability before ones with a $-$ sign, so we only need to
check the former case. For example, if $h=2g$, the instability occurs if
\begin{equation}
\nu _{n}^{2}+g\left( |R_{1}|^{2}+|R_{2}|^{2}-\sqrt{
|R_{1}|^{4}+14|R_{1}^{2}R_{2}^{2}|+|R_{2}|^{2}}\right) <0.  \label{side}
\end{equation}
Stability conditions for the plane-wave solutions to Eqs. (\ref{cnls2}) with
$V=0$ can be obtained for all the possible sign combinations of $g$ and $h$.
We do not display them here, as they are rather cumbersome to write (although
straightforward to compute).

\section{Modulated Amplitude Waves}

We now generalize the above analysis to consider the two-component GP system
in the presence of an OL potential. Toward this aim, we introduce solutions to
Eqs. (\ref{cnls2}) that describe coherent structures of the form
\begin{equation}
\psi _{j}(x,t)=R_{j}(x)\exp \left[ i(\theta (x)-\mu t)\right] \,,\quad
j=1,2\,.  \label{coher}
\end{equation}
Inserting Eq. (\ref{coher}) into Eqs. (\ref{cnls2}) and equating real and
imaginary parts of the resulting equations yields
\begin{align}
\mu R_{1}& =-R_{1}^{\prime \prime }+ R_1 \left( \theta ^{\prime }\right)
^{2}{}+gR_{1}^{3}+V(x)R_{1}+hR_{2}^{2}R_{1}+\alpha R_{2}\,,  \nonumber \\
\mu R_{2}& =-R_{2}^{\prime \prime }+ R_2 \left( \theta ^{\prime }\right)
^{2}+gR_{2}^{3}+V(x)R_{2}+hR_{1}^{2}R_{2}+\alpha R_{1}\,,  \label{R} \\
0& =R_{1}\theta ^{\prime \prime }+R_{1}^{\prime }\theta ^{\prime },
\nonumber \\
0& =R_{2}\theta ^{\prime \prime }+R_{2}^{\prime }\theta ^{\prime },
\label{theta}
\end{align}
where the prime stands for $d/dx$. Equations (\ref{theta}) imply that
\begin{equation}
\theta (x)=c_{1}\int \frac{dx^{\prime }}{R_{1}^{2}(x^{\prime })}=c_{2}\int
\frac{dx^{\prime }}{R_{2}^{2}(x^{\prime })}\,,  \label{strange}
\end{equation}
with arbitrary integration constants $c_{1}$ and $c_{2}$, so $
R_{1}(x)=bR_{2}(x)$ for some constant $b$ unless $c_{1}=c_{2}=0$.
In other words, $R_{1}(x)$ and $R_{2}(x)$ are different in this
context only when one considers solutions with null
\textquotedblleft angular momenta.\textquotedblright  \hspace{.1 cm} In the
latter situation, Eqs. (\ref{R}) assume the form
\begin{align}
R_{1}^{\prime }& =S_{1}\,,  \nonumber \\
S_{1}^{\prime }& =-\mu R_{1}+gR_{1}^{3}+hR_{1}R_{2}^{2}+\alpha
R_{2}+V(x)R_{1}\,,  \nonumber \\
R_{2}^{\prime }& =S_{2}\,,  \nonumber \\
S_{2}^{\prime }& =-\mu R_{2}+gR_{2}^{3}+hR_{1}^{2}R_{2}+\alpha
R_{1}+V(x)R_{2}\,.  \label{mat}
\end{align}
When the potential $V(x)$ is sinusoidal, Eqs. (\ref{mat}) are (linearly and
nonlinearly) coupled cubic Mathieu equations.

\section{Averaged Equations and Spatial Subharmonic Resonances}

To achieve some analytical understanding of the spatial resonances in
linearly coupled BECs, we average equations (\ref{mat}) in the physically
relevant case of the OL potential,
\begin{equation}
V(x)=V_{0}\cos (2\kappa x)\,.  \label{OL}
\end{equation}
Defining $V_{0}\equiv -\epsilon \tilde{V}_{0}$, $g\equiv \epsilon
\tilde{g}$, $h\equiv \epsilon \tilde{h}$, and $\alpha \equiv
\epsilon \tilde{\alpha}$, Eqs. (\ref{mat}) may be written
\begin{align}
R_{1}^{\prime \prime }+\mu R_{1}& =-\epsilon
\tilde{V}_{0}R_{1}\cos (2\kappa x)+\epsilon
\tilde{g}R_{1}^{3}+\epsilon \tilde{h}
R_{1}R_{2}^{2}+\epsilon \tilde{\alpha}R_{2}\,,  \nonumber \\
R_{2}^{\prime \prime }+\mu R_{2}& =-\epsilon
\tilde{V}_{0}R_{2}\cos (2\kappa x)+\epsilon
\tilde{g}R_{2}^{3}+\epsilon \tilde{h} R_{1}^{2}R_{2}+\epsilon
\tilde{\alpha}R_{1}\,.  \label{mat2}
\end{align}
Assuming $\mu >0$, we insert the ansatz
\begin{align}
R_{j}(x)& =A_{j}(x)\cos \left( \sqrt{\mu }x\right) +B_{j}(x)\sin \left(
\sqrt{\mu }x\right) \,,  \nonumber \\
R_{j}^{\prime }(x)& =-\sqrt{\mu }A_{j}(x)\sin \left(
\sqrt{\mu } x\right) +\sqrt{\mu }B_{j}(x)\cos \left(
\sqrt{\mu }x\right) \label{ansatz}
\end{align}
(with $j=1,2$) into Eqs. (\ref{mat2}). Differentiating the first equation of
(\ref{ansatz}) and comparing it with the second yields a consistency
condition,
\[
A_{j}^{\prime }\cos (\sqrt{\mu }x)+B_{j}^{\prime }\sin
(\sqrt{\mu } x)=0\,,\quad j=1,2\,,
\]
that must be satisfied for this procedure to be valid. Inserting
these equations into Eqs. (\ref{mat2}) yields a set of coupled
differential equations for $A_{j}$ and $B_{j}$, whose right-hand
sides are expanded as truncated Fourier series to isolate
contributions from different harmonics \cite{675,rand}. The
leading contribution in these equations is of $ O(\epsilon )$, so
the equations assume a general form
\begin{align}
A_{j}^{\prime }& =\epsilon F_{A_{j}}(A_{1},A_{2},B_{1},B_{2},x)+O(\epsilon
^{2})\,,  \nonumber \\
B_{j}^{\prime }& =\epsilon F_{B_{j}}(A_{1},A_{2},B_{1},B_{2},x)+O(\epsilon
^{2})\,.  \label{preavg}
\end{align}
When $\epsilon =0$, Eqs. (\ref{mat2}) decompose into two uncoupled harmonic
oscillators. We have computed the exact functions $F_{A_{j}}$ and $F_{B_{j}}$
in Eqs. (\ref{preavg}) and provide them in the Appendix.

Our objective is to isolate the parts of the functions $A_{j}(x)$
and $ B_{j}(x)$ that vary slowly in comparison with the fast
oscillations of $\cos (\sqrt{\mu }x)$ and $\sin (\sqrt{\mu
}x)$ and to derive averaged equations governing their slow
evolution. To commence averaging, we decompose $A_{j}$ and $B_{j}$
into the sum of slowly varying parts and small rapidly oscillating
ones (which are written as power-series expansions in $ \epsilon
$):
\begin{align}
A_{j}& =\bar{A}_{j}+\epsilon
W_{A_{j}}(\bar{A}_{1},\bar{A}_{2},\bar{B}_{1},
\bar{B}_{2},x)+O(\epsilon ^{2}),  \nonumber \\
B_{j}& =\bar{B}_{j}+\epsilon
W_{B_{j}}(\bar{A}_{1},\bar{A}_{2},\bar{B}_{1},
\bar{B}_{2},x)+O(\epsilon ^{2})\,.  \label{av}
\end{align}
Here, the {\it generating functions} $W_{A_{j}}$, $W_{B_{j}}$ are chosen so
as to eliminate all the rapidly oscillating terms in Eqs. (\ref{preavg})
after the substitution of Eqs. (\ref{av}).

This procedure yields evolution equations for the averaged
quantities $\bar{A}_{j}$ and $\bar{B}_{j}$\cite{675}, which we
henceforth denote simply as $ A_{j}$ and $B_{j}$.  (All other terms
in the originally defined $A_{j}$ and $ B_{j}$ are cancelled out
by the choice of the generating functions.)  As we shall see, the
slow-flow equations so derived are different in resonant and
non-resonant situations.

\subsection{The Non-Resonant Case}

When $\sqrt{\mu }\neq \kappa $ [recall that $\kappa $ is half the wave
number of the OL potential; see Eq. (\ref{OL})], which is the non-resonant
case, effective equations governing the slow evolution are
\begin{align}
A_{1}^{\prime }& =\frac{\epsilon }{\sqrt{\mu }}\left[
-\frac{3\tilde{g}}{8
}B_{1}(A_{1}^{2}+B_{1}^{2})-\frac{\tilde{\alpha}}{2}B_{2}-\frac{\tilde{h}}{4}
A_{1}A_{2}B_{2}-\frac{\tilde{h}}{8}B_{1}(A_{2}^{2}+3B_{2}^{2})\right]
+O(\epsilon ^{2})\,,  \nonumber \\
A_{2}^{\prime }& =\frac{\epsilon }{\sqrt{\mu }}\left[
-\frac{3\tilde{g}}{8
}B_{2}(A_{2}^{2}+B_{2}^{2})-\frac{\tilde{\alpha}}{2}B_{1}-\frac{\tilde{h}}{4}
A_{1}A_{2}B_{1}-\frac{\tilde{h}}{8}B_{2}(A_{1}^{2}+3B_{1}^{2})\right]
+O(\epsilon ^{2})\,,  \nonumber \\
B_{1}^{\prime }& =\frac{\epsilon }{\sqrt{\mu }}\left[
\frac{3\tilde{g}}{8}
A_{1}(A_{1}^{2}+B_{1}^{2})+\frac{\tilde{\alpha}}{2}A_{2}+\frac{\tilde{h}}{4}
A_{2}B_{1}B_{2}+\frac{\tilde{h}}{8}A_{1}(3A_{2}^{2}+B_{2}^{2})\right]
+O(\epsilon ^{2})\,,  \nonumber \\
B_{2}^{\prime }& =\frac{\epsilon }{\sqrt{\mu }}\left[
\frac{3\tilde{g}}{8}
A_{2}(A_{2}^{2}+B_{2}^{2})+\frac{\tilde{\alpha}}{2}A_{1}+\frac{\tilde{h}}{4}
A_{1}B_{1}B_{2}+\frac{\tilde{h}}{8}A_{2}(3A_{1}^{2}+B_{1}^{2})\right]
+O(\epsilon ^{2})\,.  \label{nonres}
\end{align}
In this case, the OL does not contribute to $O(\epsilon )$ terms, so the
terms explicitly written in Eqs. (\ref{nonres}) correspond to what one would
obtain from coupled Duffing equations, as Eqs. (\ref{mat2}) reduce to
coupled Duffing oscillators in the absence of the OL potential \cite{param}.
These contributions yield the wavenumber-amplitude relations for decoupled
condensates,\cite{mapbecprl,mapbec} as well as mode-wavenumber relations
produced by the coupling terms \cite{675}.

The non-resonant equations (\ref{nonres}) give rise to three types
of equilibria, at which $A_{1}^{\prime }=A_{2}^{\prime
}=B_{1}^{\prime }=B_{2}^{\prime }=0$: the trivial (zero)
equilibrium and those which we will call double modes and
quadruple modes. These have, respectively, two and four nonzero
amplitudes $A_{j}$, $B_{j}$. Single-mode and triple-mode
equilibria do not exist. Different double modes that can be found
are $\pi /2 $ phase shifts of each other: these are
\textquotedblleft $A_{1}A_{2}$\textquotedblright\ equilibria
with $A_{1},\,A_{2}\neq 0$ and $ B_{1}=B_{2}=0 $, and
\textquotedblleft $B_{1}B_{2}$\textquotedblright\ ones with
$A_{1}=A_{2}=0$ and $B_{1},\,B_{2}\neq 0$.

The $A_{1}A_{2}$ equilibria satisfy
\begin{equation}
A_{1}^{2}=A_{2}^{2}=\mp \frac{4\alpha }{3(g+h)}\,,  \label{eqref1}
\end{equation}
where the signs $-$ and $+$ arise, respectively, when $(g+h)<0$,
and $ (g+h)>0 $ (recall that $\alpha >0$). In the former and
latter cases, we find that $A_{1}=A_{2}$ and $A_{1}=-A_{2}$,
respectively. This yields the following two $A_{1}A_{2}$
equilibria:
\begin{align}
(A_{1},A_{2},B_{1},B_{2})& =\pm \left( \sqrt{\frac{4\alpha
}{3(g+h)}},-\sqrt{
\frac{4\alpha }{3(g+h)}},0,0\right) ,\quad {\rm if}~~g+h>0\,;  \nonumber \\
(A_{1},A_{2},B_{1},B_{2})& =\pm \left( \sqrt{\frac{-4\alpha
}{3(g+h)}},\sqrt{ \frac{-4\alpha }{3(g+h)}},0,0\right) ,\quad {\rm
if}~~g+h<0\,.  \label{aa}
\end{align}
Similar expressions for the $B_{1}B_{2}$ equilibria are obtained
by phase-shifting the $A_{1}A_{2}$ modes by $\pi /2$.

We have examined the stability of the approximate stationary solutions
corresponding to the double-mode equilibria obtained above with direct
simulations of the coupled GP equations (\ref{cnls2}). Typically, the
simulations generate solutions that oscillate in time (as the initial
configurations are not exact stationary solutions) without
developing any apparent instability.

One can also find four sets of quadruple-mode equilibria in which
$ A_{1}^{2}=A_{2}^{2}$ and $B_{1}^{2}=B_{2}^{2}$. In the first two
sets, $ A_{2} $ is arbitrary:
\begin{align}
(A_{1},A_{2},B_{1},B_{2})& =\pm \left( -A_{2},A_{2},\pm
\sqrt{-A_{2}^{2}+ \frac{4\alpha }{3(g+h)}},\mp
\sqrt{-A_{2}^{2}+\frac{4\alpha }{3(g+h)}}\right)
,~{\rm if}~~g+h>0\,,  \nonumber \\
(A_{1},A_{2},B_{1},B_{2})& =\pm \left( A_{2},A_{2},\pm
\sqrt{-A_{2}^{2}- \frac{4\alpha }{3(g+h)}},\pm
\sqrt{-A_{2}^{2}-\frac{4\alpha }{3(g+h)}}\right) ,~{\rm
if}~~g+h<0\,.  \label{quad1}
\end{align}
In the second two sets, $B_{2}$ is arbitrary:
\begin{align}
(A_{1},A_{2},B_{1},B_{2})& =\pm \left( \pm
\sqrt{-B_{2}^{2}+\frac{4\alpha } {3(g+h)}},\mp
\sqrt{-B_{2}^{2}+\frac{4\alpha }{3(g+h)}},-B_{2},B_{2}\right) ,~
{\rm if}~~g+h>0\,,  \nonumber \\
(A_{1},A_{2},B_{1},B_{2})& =\pm \left( \pm
\sqrt{-B_{2}^{2}-\frac{4\alpha } {3(g+h)}},\pm
\sqrt{-B_{2}^{2}-\frac{4\alpha }{3(g+h)}},B_{2},B_{2}\right) ,~
{\rm if}~~g+h<0\,.  \label{quad2}
\end{align}
Each of the expressions (\ref{quad1}) and (\ref{quad2}) includes four
equilibria, as there are two possible choices of the exterior signs. The
presence of the arbitrary amplitudes in these expressions means that the
quadruple-mode stationary solutions are obtained as rotations
 of the above double-mode ones given, respectively, by Eqs. (\ref{aa}) and by
 those same equations with an additional $\pi /2$ phase shift. Accordingly, 
direct simulations of Eqs. (\ref{cnls2}) starting with the approximate
quadruple-mode stationary states reveal only oscillations but no instability
growth, just as with simulations initiated by the approximate dual-mode
stationary solutions.

\subsection{Subharmonic Resonances}

The most fundamental spatial resonance is a subharmonic one, of
type $ 2\!:\!1\!:\!1$\cite{675,rand,gucken}. In this
situation, the parameter $ \mu $ from the initial plane-wave
approximation (\ref{plane}) [recall that $\mu _{1}=\mu
_{2}\equiv \mu $] is of the form
\begin{equation}
\mu =\kappa ^{2}+\epsilon \tilde{\mu}_{1}+O(\epsilon ^{2})\,,
\label{detune}
\end{equation}
where $\epsilon \tilde{\mu}_{1}$ is the {\it detuning}
constant\cite {rand,675,param}.  [Recall that $\epsilon $ is a small
parameter; we assume $ \tilde{\mu}_{1}=O(1)$.]  In this
situation, new terms occur in Eqs. (\ref {nonres}). This leads to
equations that include a contribution from the OL potential,
\begin{align}
A_{1}^{\prime }& =\frac{\epsilon }{\kappa }\left[ \left(
\frac{\tilde{\mu} _{1}}{2}-\frac{\tilde{V}_{0}}{4}\right)
B_{1}-\frac{3\tilde{g}}{8}
B_{1}(A_{1}^{2}+B_{1}^{2})-\frac{\tilde{\alpha}}{2}B_{2}-\frac{\tilde{h}}{4}
A_{1}A_{2}B_{2}-\frac{\tilde{h}}{8}B_{1}(A_{2}^{2}+3B_{2}^{2})\right]
+O(\epsilon ^{2})\,,  \nonumber \\
A_{2}^{\prime }& =\frac{\epsilon }{\kappa }\left[ \left(
\frac{\tilde{\mu} _{1}}{2}-\frac{\tilde{V}_{0}}{4}\right)
B_{2}-\frac{3\tilde{g}}{8}
B_{2}(A_{2}^{2}+B_{2}^{2})-\frac{\tilde{\alpha}}{2}B_{1}-\frac{\tilde{h}}{4}
A_{1}A_{2}B_{1}-\frac{\tilde{h}}{8}B_{2}(A_{1}^{2}+3B_{1}^{2})\right]
+O(\epsilon ^{2})\,,  \nonumber \\
B_{1}^{\prime }& =\frac{\epsilon }{\kappa }\left[ -\left(
\frac{\tilde{\mu }_{1}}{2}+\frac{\tilde{V}_{0}}{4}\right)
A_{1}+\frac{3\tilde{g}}{8}
A_{1}(A_{1}^{2}+B_{1}^{2})+\frac{\tilde{\alpha}}{2}A_{2}+\frac{\tilde{h}}{4}
A_{2}B_{1}B_{2}+\frac{\tilde{h}}{8}A_{1}(3A_{2}^{2}+B_{2}^{2})\right]
+O(\epsilon ^{2})\,,  \nonumber \\
B_{2}^{\prime }& =\frac{\epsilon }{\kappa }\left[ -\left(
\frac{\tilde{\mu }_{1}}{2}+\frac{\tilde{V}_{0}}{4}\right)
A_{2}+\frac{3\tilde{g}}{8}
A_{2}(A_{2}^{2}+B_{2}^{2})+\frac{\tilde{\alpha}}{2}A_{1}+\frac{\tilde{h}}{4}
A_{1}B_{1}B_{2}+\frac{\tilde{h}}{8}A_{2}(3A_{1}^{2}+B_{1}^{2})\right]
+O(\epsilon ^{2})\,.  \label{res}
\end{align}

Equations (\ref{res}) have three types of equilibria when $\alpha \neq 0$:
the trivial one, double modes, and quadruple modes. When $\alpha =0$, we
also find single-mode equilibria and extra double-mode ones. Triple-mode
stationary solutions never appear. All the equilibria of Eqs. (\ref{res}),
except for the trivial one, correspond to spatially periodic stationary
solutions of the underlying system (\ref{mat2}).

\begin{figure}[tbp]
{\epsfig{file=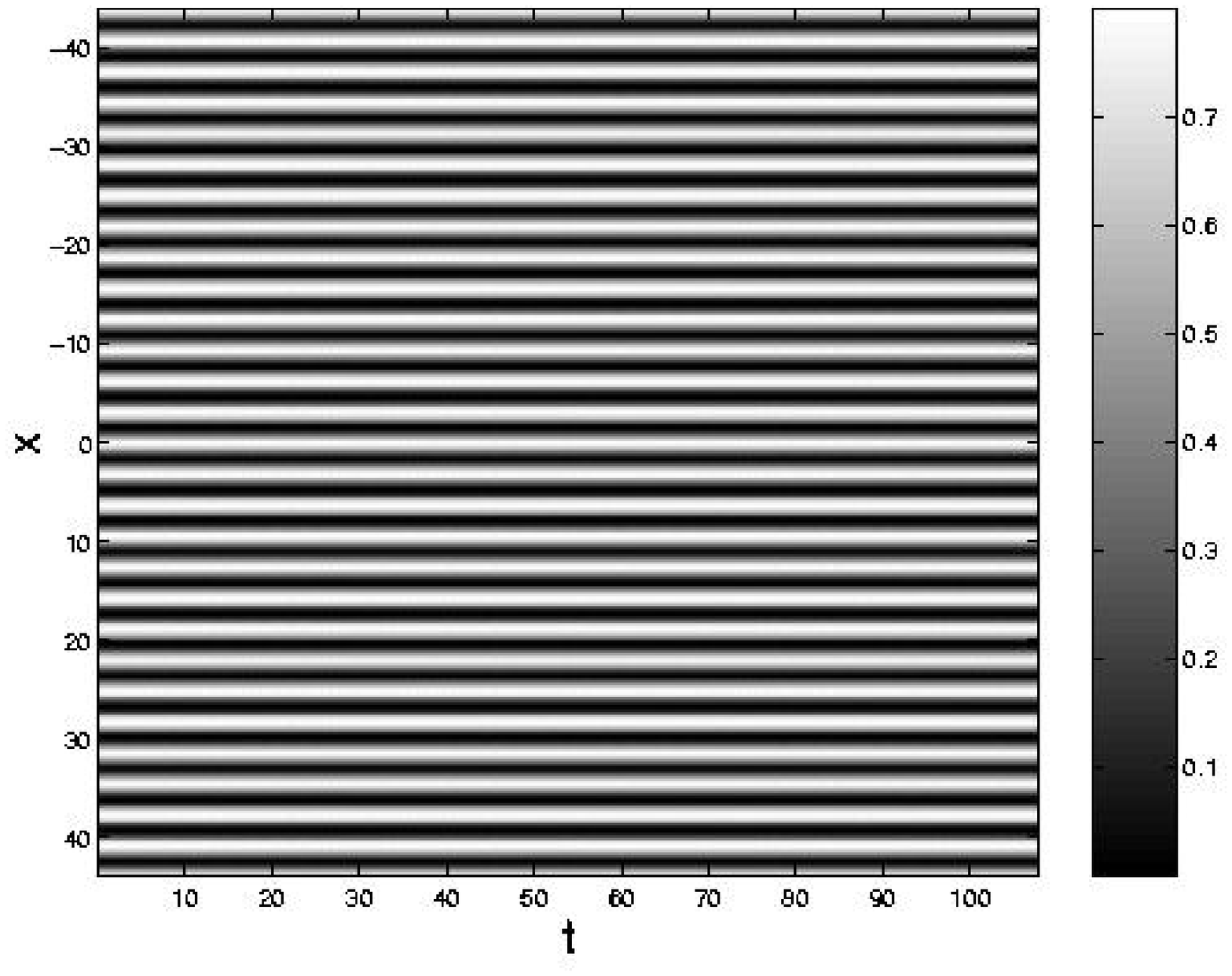, width=7.5cm,angle=0, clip=}}
{\epsfig{file=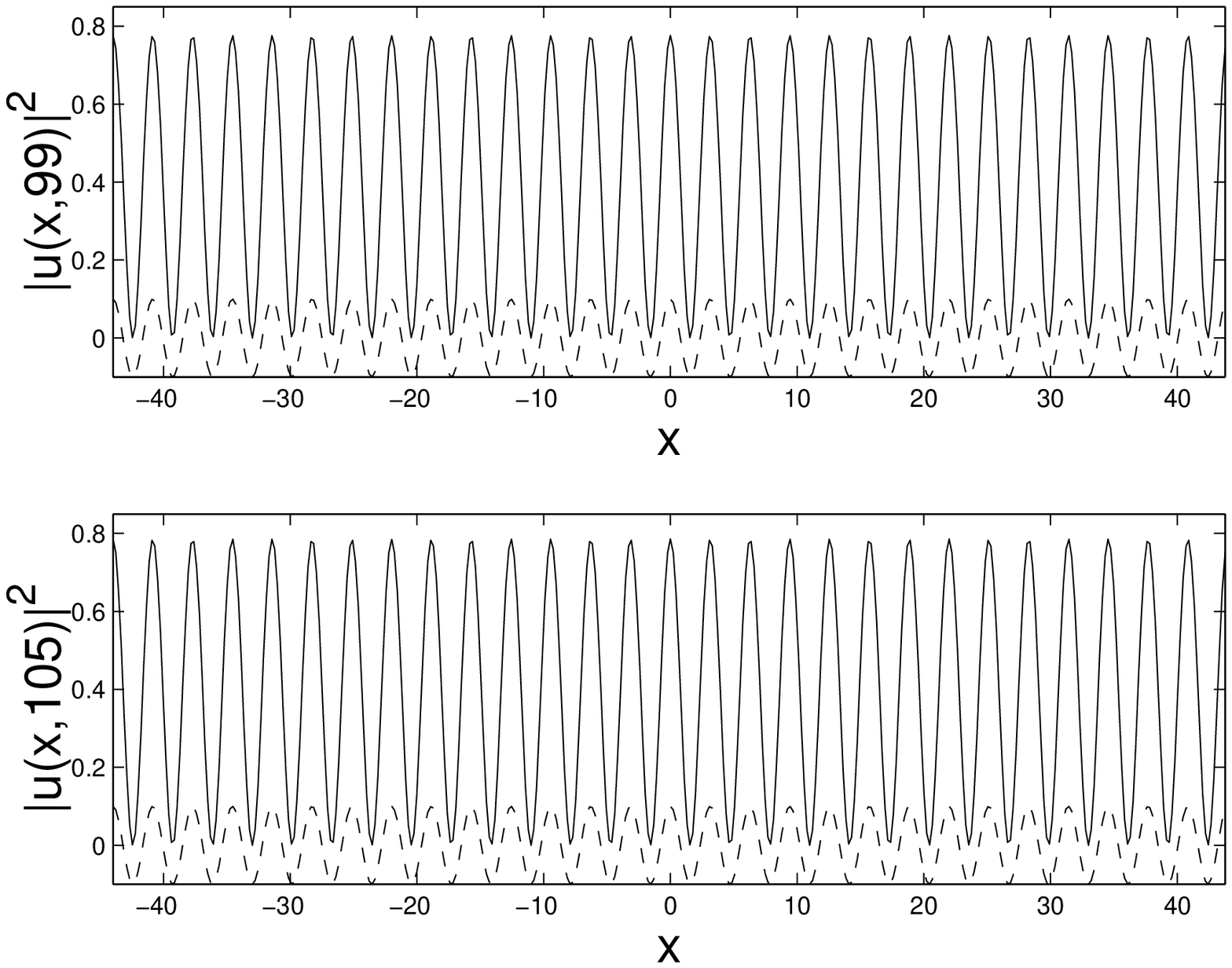, width=7.5cm,angle=0, clip=}}
\caption{An
example of evolution of the $A_{1}A_{2}$ double mode with {\it
equal} amplitudes $\left\vert A_{1}\right\vert $ and $\left\vert
A_{2}\right\vert $ in the case of the $2\!:\!1\!:\!1$ subharmonic
resonance, constructed as per Eq. (\protect\ref{a1a2type1}) for
$\protect\kappa = \protect\mu =1$, $V_{0}=0.1$, $g=h=0.025$,
and $\protect\alpha =0.02$. The left subplot shows the
spatio-temporal evolution of $\left\vert \psi _{1}\right\vert
^{2}$ by means of grayscale contour plots ($\left\vert
\protect\psi _{2}\right\vert ^{2}$ behaves similarly). The right
subplot displays snapshots of the field $\left\vert \protect\psi
_{1}\right\vert ^{2} $ for $t=99$ (upper panel) and $t=105$ (lower
panel). In these panels, the optical-lattice potential is
also shown by a dashed line. The results have been obtained
through numerical integration of Eqs. (\protect \ref{cnls2}) in
time.} \label{mpfig3}
\end{figure}

There are two kinds of $A_{1}A_{2}$ (double-mode) equilibria. The first
satisfies $A_{1}^{2}=A_{2}^{2}$, so that the two components have equal
amplitudes:
\begin{align}
(A_{1},A_{2},B_{1},B_{2})& =\pm \left( \sqrt{\frac{-4\alpha +2(2\mu
_{1}+V_{0})}{3(g+h)}},\sqrt{\frac{-4\alpha +2(2\mu _{1}+V_{0})}{3(g+h)}} ,0,0\right) \,,  \nonumber \\
(A_{1},A_{2},B_{1},B_{2})& =\pm \left( \sqrt{\frac{4\alpha
+2(2\mu _{1}+V_{0})}{3(g+h)}},-\sqrt{\frac{4\alpha +2(2\mu
_{1}+V_{0})}{3(g+h)}} ,0,0\right) \,.  \label{a1a2type1}
\end{align}
A crucial issue is the dynamical stability of
these solutions, which we tested with direct simulations of the
underlying equations (\ref{cnls2}). We found that they are {\em
stable}, as exemplified in Fig. \ref{mpfig3} for $\kappa =\mu
=1$.

The other $A_{1}A_{2}$ double-mode equilibrium has
{\it unequal} components $A_{1}$ and $A_{2}$ [note that in the
non-resonant case considered above, the double-mode equilibria,
which are given by Eqs. (\ref {aa}) and by a $\pi /2$ phase shift
thereof, always have equal nonzero components]:
\[
A_{1}^{2}+A_{2}^{2}=\frac{2(2\mu _{1}+V_{0})}{3g}>0\,,
\]
\begin{align}
(A_{1},A_{2},B_{1},B_{2})& =\pm \left( \sqrt{\frac{2\mu
_{1}+V_{0}}{3g} \pm \frac{1}{3}\sqrt{\frac{(2\mu
_{1}+V_{0})^{2}}{g^{2}}-\frac{16\alpha
^{2}}{(g-h)^{2}}}}\,,\right.  \nonumber \\
& \qquad \left. \sqrt{\frac{2\mu _{1}+V_{0}}{3g}\mp
\frac{1}{3}\sqrt{ \frac{(2\mu
_{1}+V_{0})^{2}}{g^{2}}-\frac{16\alpha ^{2}}{(g-h)^{2}}}}
,0,0\right) \,,  \label{unequal}
\end{align}
where the interior $+$ sign in the first component corresponds to the $-$
sign in the second, and vice versa. The exterior sign $\pm $ is independent
of the interior one. A necessary condition for the existence of this
solution is
\[
\left\vert \frac{2\mu _{1}+V_{0}}{g}\right\vert \geq
\frac{4\alpha }{|g-h| }\,.
\]
In particular, when $h=2g$, which is a case of special physical relevance
(as explained above), the solution becomes
\begin{align}
(A_{1},A_{2},B_{1},B_{2})& =\pm \left( \sqrt{\frac{1}{3g}\left[
2\mu _{1}+V_{0}\pm \sqrt{(2\mu _{1}+V_{0})^{2}-16\alpha
^{2}}\right] }
\,,\right. \\
& \qquad \left. \sqrt{\frac{1}{3g}\left[ 2\mu _{1}+V_{0}\mp
\sqrt{ (2\mu _{1}+V_{0})^{2}-16\alpha ^{2}}\right] },0,0\right)
\,,
\end{align}
provided $|2\mu _{1}+V_{0}|\geq 4\alpha $.

In fact, the existence of pairs of equilibria in which the two
components have unequal amplitudes that are mirror images of each
other is a manifestation of {\it spontaneous symmetry breaking} in
the present model, which is described by the symmetric system of
coupled equations (\ref{cnls2}). A similar phenomenon was studied
in detail (in terms of soliton solutions) in the aforementioned
model of dual-core nonlinear optical fibers, which includes only
linear coupling between two equations \cite {UNSW}.

\begin{figure}[tbp]
{\epsfig{file=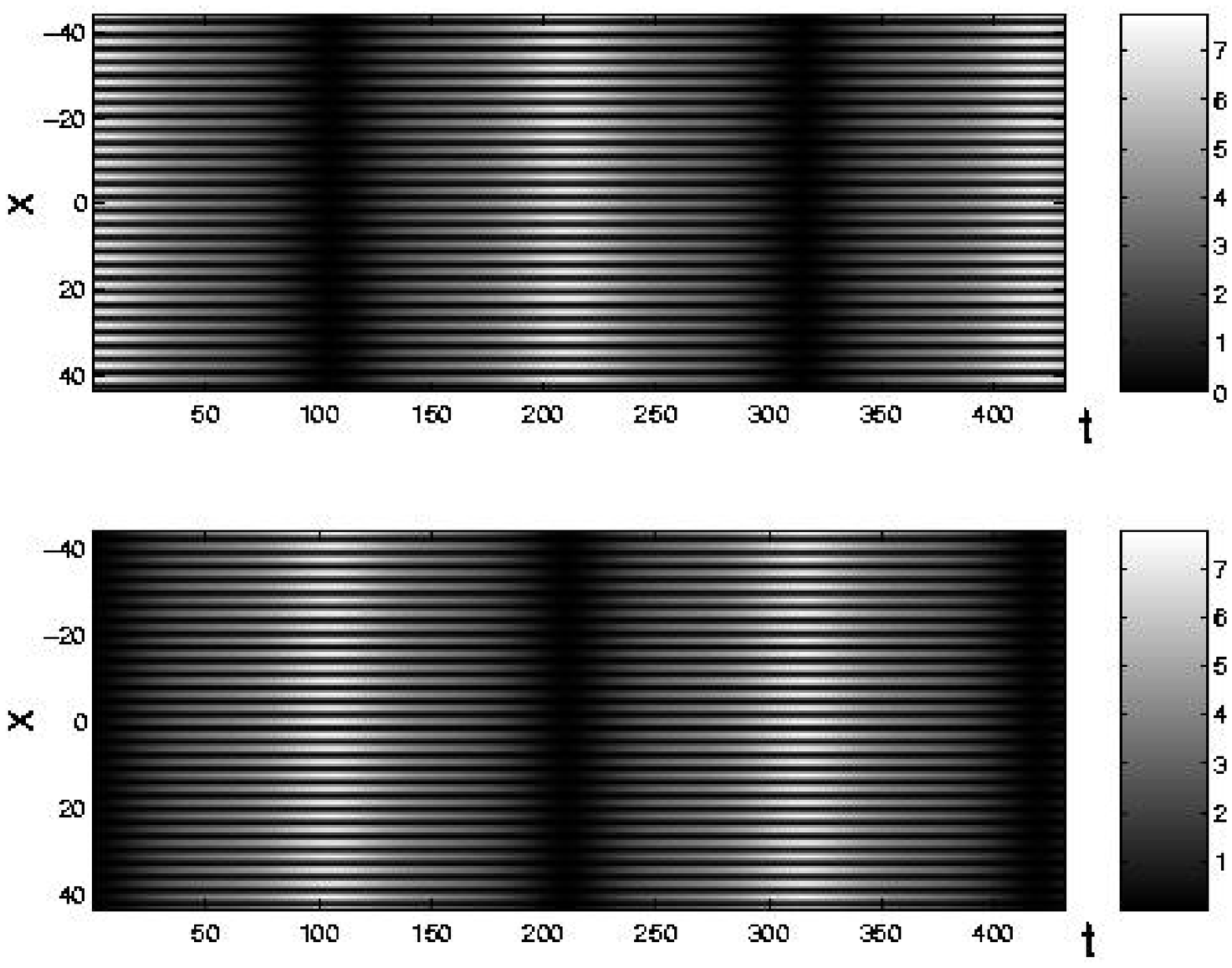, width=7.5cm,angle=0, clip=}}
{\epsfig{file=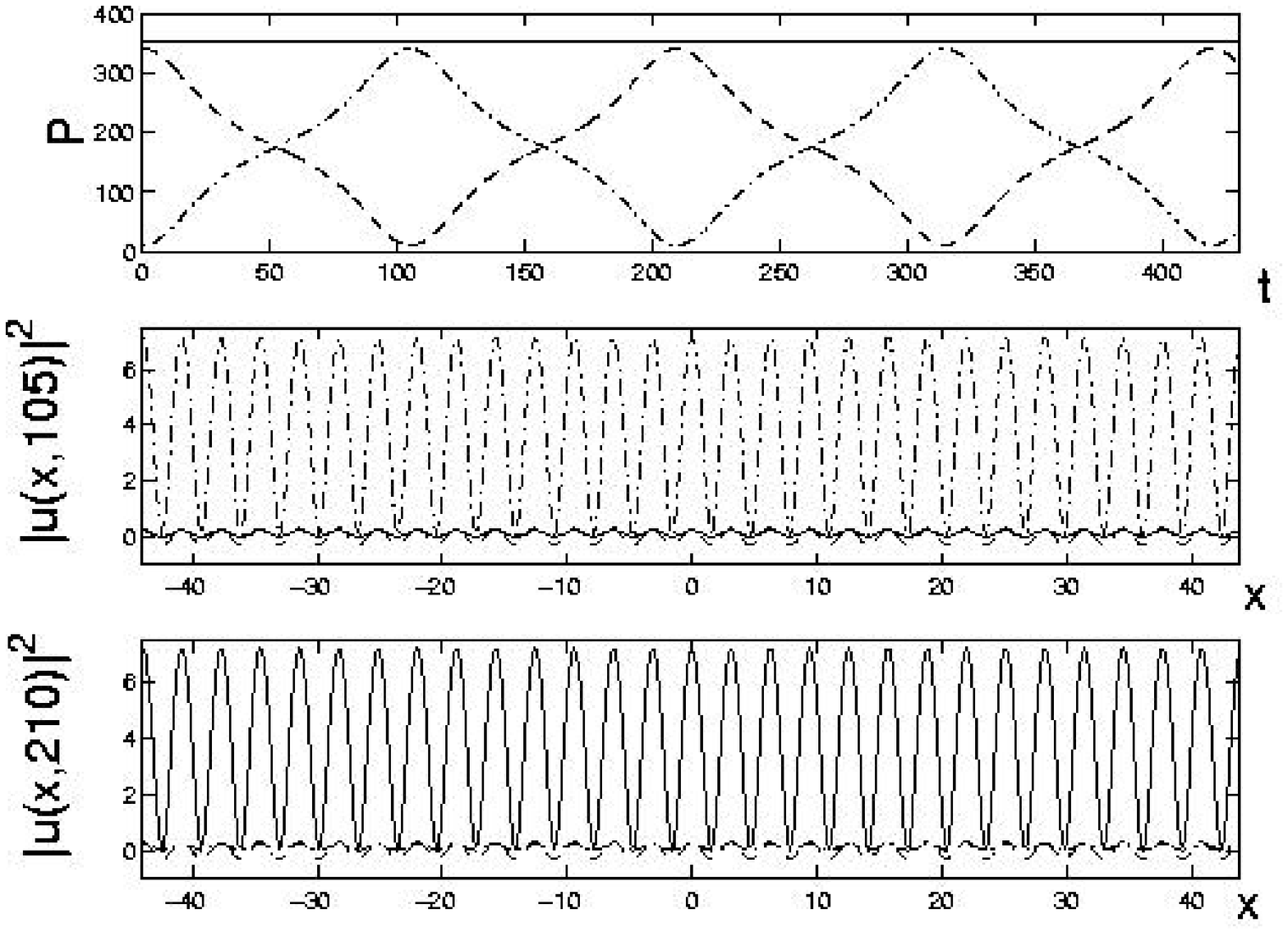, width=7.5cm,angle=0, clip=}}
\caption{The same as in Fig. \protect\ref{mpfig3}, but for the
$A_{1}A_{2}$ double mode with {\it unequal} amplitudes $\left\vert
A_{1}\right\vert $ and $\left\vert A_{2}\right\vert $, given by
Eq. (\protect\ref{unequal}). The top panel in the right subplot
shows the time evolution of the numbers of atoms in the two
components, $N_{1,2}=\protect\int |\psi_{1,2}|^{2}dx$, by dashed and
dash-dotted lines (the sum of the two, $N=N_{1}+N_{2}$, is shown
by the solid line). The right panel also shows the fields
$\left\vert \protect\psi _{1}\right\vert ^{2}$ and $\left\vert
\protect\psi_{2}\right\vert ^{2}$ (by solid and dash-dotted lines,
respectively) at $ t=105$ and $t=210$. The OL potential is shown
by the dashed line. Oscillations of matter between the two
components are clearly discernible. The parameters are $g=0.025$,
$h=0.005$, $\protect\alpha =-0.02$, $V_{0}=0.3$, and
$\protect\kappa =\protect\mu =1$.  (Recall that the sign of $\alpha$ 
can be chosen arbitrarily.)} \label{mpfig4}
\end{figure}

The stability of the asymmetric stationary solutions, which
correspond to equilibria with unequal components, was also
simulated in the framework of Eqs. (\ref{cnls2}).  We show the results
of a typical simulation in Fig. \ref {mpfig4}. As seen in
the figure, these states are subject to
periodic oscillations between the two
components (which is possible only in the presence of
the linear coupling between them).

The resonant equations (\ref{res}) give rise to two types of $B_{1}B_{2}$
dual-mode equilibria. The first satisfies $B_{1}^{2}=B_{2}^{2}$ and
\begin{align}
(A_{1},A_{2},B_{1},B_{2})& =\pm \left( 0,0,\sqrt{\frac{-4\alpha +2(2\mu
_{1}-V_{0})}{3(g+h)}},\sqrt{\frac{-4\alpha +2(2\mu _{1}-V_{0})}{3(g+h)}} \right) \,,  \nonumber \\
(A_{1},A_{2},B_{1},B_{2})& =\pm \left( 0,0,\sqrt{\frac{4\alpha
+2(2\mu _{1}-V_{0})}{3(g+h)}},-\sqrt{\frac{4\alpha +2(2\mu
_{1}-V_{0})}{3(g+h)}} \right) \,.  \label{b1b2type1}
\end{align}
The second type satisfies
\[
B_{1}^{2}+B_{2}^{2}=\frac{2(2\mu _{1}-V_{0})}{3g}>0\,,
\]
\begin{align}
(A_{1},A_{2},B_{1},B_{2})& =\pm \left( 0,0,\sqrt{\frac{2\mu
_{1}-V_{0}} {3g}\pm \frac{1}{3}\sqrt{\frac{(2\mu
_{1}-V_{0})^{2}}{g^{2}}-\frac{16\alpha
^{2}}{(g-h)^{2}}}},\right.  \nonumber \\
& \qquad \left. \sqrt{\frac{2\mu _{1}-V_{0}}{3g}\mp
\frac{1}{3}\sqrt{ \frac{(2\mu
_{1}-V_{0})^{2}}{g^{2}}-\frac{16\alpha ^{2}}{(g-h)^{2}}}} \right)
\,.  \label{unequalBB}
\end{align}
As above, the interior $+$ sign in the first component is paired to the $-$
sign in the second, and vice versa, whereas the exterior $\pm $ is
independent. A necessary condition for the existence of this solution is
\begin{equation}
\left\vert \frac{2\mu _{1}-V_{0}}{g}\right\vert \geq
\frac{4\alpha }{|g-h| }\,.  \label{cond}
\end{equation}
When $h=2g$, the present solution becomes
\begin{align}
(A_{1},A_{2},B_{1},B_{2})& =\pm \left( 0,0,\sqrt{\frac{1}{3g}\left[ 2\mu
_{1}-V_{0}\pm \sqrt{(2\mu _{1}-V_{0})^{2}-16\alpha ^{2}}\right] },\right.
\\
& \qquad \left. \sqrt{\frac{1}{3g}\left[ 2\mu _{1}-V_{0}\mp
\sqrt{ (2\mu _{1}-V_{0})^{2}-16\alpha ^{2}}\right] }\right) \,,
\end{align}
provided $|2\mu _{1}-V_{0}|\geq 4\alpha $.

Unlike the non-resonant case, the resonant $B_{1}B_{2}$ modes are {\em
not} precise phase shifts of the $A_{1}A_{2}$ modes,
as the
spatial parametric excitation resulting from the OL has only the
cosine harmonic.  Nevertheless, the equations describing these two
classes of modes are similar, differing only in the sign of $V_{0}$.
Direct simulations demonstrate that the stability of stationary solutions
corresponding to the $B_{1}B_{2}$ equilibria is the same as in the
case of the $A_{1}A_{2}$ double-mode equilibria considered above:
the symmetric ones with $\left\vert B_{1}\right\vert =\left\vert
B_{2}\right\vert $ are {\it stable}, and the asymmetric solutions
with $\left\vert B_{1}\right\vert \neq \left\vert B_{2}\right\vert
$ are {\it unstable}.

We have also found two sets of quadruple modes in the resonant case. The
first set satisfies $B_{1}=B_{2}$, $A_{1}=-A_{2}$, and
\begin{align}
A_{1}^{2}& =\frac{V_{0}}{2h}+\frac{\alpha }{h}+\frac{2\mu _{1}}{3g+h}, \\
B_{1}^{2}& =-\frac{V_{0}}{2h}-\frac{\alpha }{h}+\frac{2\mu _{1}}{3g+h}\,.
\end{align}
A necessary condition for its existence is
\[
\frac{2\mu _{1}}{3g+h}>\left\vert \frac{V_{0}}{2h}+\frac{\alpha
}{h} \right\vert \,,
\]
and hence it is necessary that $\mu _{1}/(3g+h)>0$. The second set of
quadruple modes satisfies $B_{1}=-B_{2}$, $A_{1}=A_{2}$, and
\begin{align}
A_{1}^{2}& =\frac{V_{0}}{2h}-\frac{\alpha }{h}+\frac{2\mu _{1}}{3g+h}\,,
\\
B_{1}^{2}& =-\frac{V_{0}}{2h}+\frac{\alpha }{h}+\frac{2\mu _{1}}{3g+h}\,.
\end{align}
A necessary existence condition for this mode to exist is
\[
\frac{2\mu _{1}}{3g+h}>\left\vert \frac{V_{0}}{2h}-\frac{\alpha
}{h} \right\vert \,,
\]
which also implies that $\mu _{1}/(3g+h)>0$.

\begin{figure}[tbp]
\centerline{ {\epsfig{file=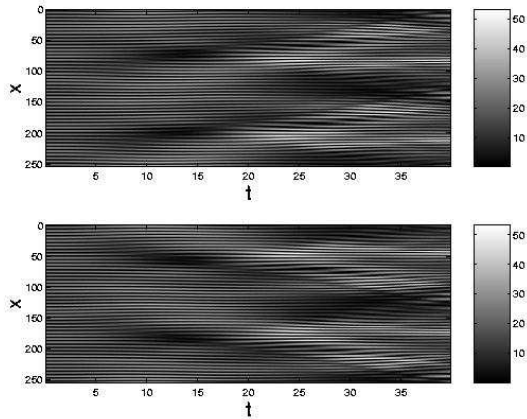,width=7.5cm,angle=0, clip=}}}
\caption{A typical example of the long-wave instability of a
quadruple (four-amplitude) stationary state in the resonant case.
The parameters are $ \protect\mu =1.5791$, $\protect\kappa
=1.2320$, $V_{0}=0.1$, $g=0.025$, $ h=0.05$ and $\protect\alpha
=0$, and the size of the integration box is $ L=255$.}
\label{mpfig5}
\end{figure}

We considered quadruple modes in the presence of detuning, so
$ \mu _{1}\neq 0$. This is rather difficult to implement
numerically, as---in view of the periodic boundary conditions in
$x$ employed in the numerical integration scheme---it is necessary
to match {\it both} the potential and initial condition to the
size of the integration domain. Nevertheless, we were able to
perform stability simulations in this case too.  We show an example of
these simulations in Fig. \ref{mpfig5}. We observe that the quadruple
mode is {\it unstable} against long-wave perturbations, even if
the simulations are run with $\alpha =0$ (no linear coupling). We
have not observed stable quadruple states.

When $\alpha =0$ (no linear coupling between BEC components), one
can find additional double-mode equilibria and four single-mode
ones, the latter of which take the form $(A_{1},0,0,0)$,
$(0,A_{2},0,0)$, $(0,0,B_{1},0)$, $ (0,0,0,B_{2})$, with
\begin{align}
A_{1}^{2}& =A_{2}^{2}=\frac{2(2\mu _{1}+V_{0})}{3g}>0\,, \\
B_{1}^{2}& =B_{2}^{2}=-\frac{2(2\mu _{1}-V_{0})}{3g}>0\,.
\end{align}
The $A_{j}$- and $B_{j}$-modes both exist when $V_{0}/g>0$. In
this case ($\alpha =0$), matter cannot be exchanged between the
components. In this same situation, there is also an $A_{1}B_{2}$
double-mode equilibrium [of the form $ (A_{1},0,0,B_{2})$], which
satisfies
\begin{align}
A_{1}^{2}& =\frac{4\mu _{1}}{3g+h}+\frac{2V_{0}}{3g-h}>0\,,  \nonumber \\
B_{2}^{2}& =\frac{4\mu _{1}}{3g+h}-\frac{2V_{0}}{3g-h}>0\,.  \label{a1b2}
\end{align}
From Eq. (\ref{a1b2}), it follows that a necessary condition for
this mode to exist is $8\mu _{1}/(3g+h)>0$. Its counterpart is
an $A_{2}B_{1}$ equilibrium, in which the subscripts $1$ and $2$
are swapped in Eq. (\ref {a1b2}).

One can extend the analysis to higher-order spatial resonances in BECs (from
the lowest subharmonic resonance considered here) either by considering
higher-order corrections to the averaged equations, or by employing a
perturbative scheme based on elliptic functions, as has been done for
single-component BECs in OLs\cite{mapbec,mapbecprl}. Toward this aim, it may
be fruitful to utilize an action-angle formulation and the elliptic-function
structure of solutions to Eqs. (\ref{mat}) when $V_{0}=0$. However, detailed
consideration of higher-order resonances is beyond the scope of this work.

\section{Ternary BECs in Optical Lattices}

To evince the generality of the above analysis, we briefly
consider its extension to a BEC model of three hyperfine states
coupled by two different microwave fields, which is
also a physically relevant situation \cite{pengels}. The corresponding
coupled GP equations (with $\hbar =1$ and $ m=1/2$) are
\begin{align}
i\frac{\partial \psi _{1}}{\partial t}& =-\nabla ^{2}\psi _{1}+g|\psi
_{1}|^{2}\psi _{1}+V(x)\psi _{1}+h_{12}|\psi _{2}|^{2}\psi _{1}+h_{13}|\psi
_{3}|^{2}\psi _{1}+\alpha _{12}\psi _{2}+\alpha _{13}\psi _{3}\,,  \nonumber
\\
i\frac{\partial \psi _{2}}{\partial t}& =-\nabla ^{2}\psi _{2}+g|\psi
_{2}|^{2}\psi _{2}+V(x)\psi _{2}+h_{12}|\psi _{1}|^{2}\psi _{2}+h_{23}|\psi
_{3}|^{2}\psi _{2}+\alpha _{12}\psi _{1}+\alpha _{23}\psi _{3}\,,  \nonumber
\\
i\frac{\partial \psi _{3}}{\partial t}& =-\nabla ^{2}\psi _{3}+g|\psi
_{3}|^{3}\psi _{3}+V(x)\psi _{3}+h_{13}|\psi _{1}|^{2}\psi _{3}+h_{23}|\psi
_{2}|^{2}\psi _{3}+\alpha _{13}\psi _{1}+\alpha _{23}\psi _{2}\,,
\label{cnls3}
\end{align}
where the self- and cross-scattering coefficients are
$g := g_{1} = g_{2} = g_{3}$ and $h_{jk}$, and the linear coupling constants
 are $\alpha _{jk}$.

As in the binary case, we start with the general form
(\ref{coher}) for stationary solutions, with $\theta
_{1}(x)=\theta _{2}(x)=\theta _{3}(x)\equiv \theta (x)$ and
$\mu _{1}=\mu _{2}=\mu _{3}\equiv \mu $. Then, as done above,
we set $c_{j}=0$ (i.e., $\theta =0$) in Eqs. (\ref{strange}) to consider
standing wave solutions and arrive at the following equations:
\begin{align}
R_{1}^{\prime }& =S_{1}\,,  \nonumber \\
S_{1}^{\prime }& =-\mu
R_{1}+gR_{1}^{3}+h_{12}R_{1}R_{2}^{2}+h_{13}R_{1}R_{3}^{2}+\alpha
_{12}R_{2}+\alpha _{13}R_{3}+V(x)R_{1}\,,  \nonumber \\
R_{2}^{\prime }& =S_{2}\,,  \nonumber \\
S_{2}^{\prime }& =-\mu
R_{2}+gR_{2}^{3}+h_{12}R_{1}^{2}R_{2}+h_{23}R_{2}R_{3}^{2}+\alpha
_{12}R_{1}+\alpha _{23}R_{3}+V(x)R_{2}\,,  \nonumber \\
R_{3}^{\prime }& =S_{3}\,,  \nonumber \\
S_{3}^{\prime }& =-\mu
R_{3}+gR_{3}^{3}+h_{13}R_{1}^{2}R_{3}+h_{23}R_{2}^{2}R_{3}+\alpha
_{13}R_{1}+\alpha _{23}R_{2}+V(x)R_{3}\,,  \label{mat3}
\end{align}
where $V(x)$ is the sinusoidal OL potential, as before.

One can average Eqs. (\ref{mat3}) with the same procedure that we applied to
Eqs. (\ref{mat}) and thereby derive both resonant and non-resonant equations
describing the system's slow dynamics. In particular, for the most
fundamental resonant case (the lowest-order, $2\!:\!1\!:\!1\!:\!1$,
resonance), the averaged equations are
\begin{align}
A_{1}^{\prime }& =\frac{\epsilon }{\kappa }\left[ \left(
\frac{\tilde{\mu} _{1}}{2}-\frac{\tilde{V}_{0}}{4}\right)
B_{1}-\frac{3\tilde{g}}{8}
B_{1}(A_{1}^{2}+B_{1}^{2})-\frac{\tilde{\alpha}_{12}}{2}B_{2}-\frac{\tilde{
\alpha}_{13}}{2}B_{3}-\frac{\tilde{h}_{12}}{4}A_{1}A_{2}B_{2}-\frac{\tilde{h} _{13}}{4}A_{1}A_{3}B_{3}\right.  \nonumber \\
& \qquad \left.
-\frac{\tilde{h}_{12}}{8}B_{1}(A_{2}^{2}+3B_{2}^{2})-\frac{
\tilde{h}_{13}}{8}B_{1}(A_{3}^{2}+3B_{3}^{2})\right] +O(\epsilon
^{2})\,,
\nonumber \\
A_{2}^{\prime }& =\frac{\epsilon }{\kappa }\left[ \left(
\frac{\tilde{\mu} _{1}}{2}-\frac{\tilde{V}_{0}}{4}\right)
B_{2}-\frac{3\tilde{g}}{8}
B_{2}(A_{2}^{2}+B_{2}^{2})-\frac{\tilde{\alpha}_{12}}{2}B_{1}-\frac{\tilde{
\alpha}_{23}}{2}B_{3}-\frac{\tilde{h}_{12}}{4}A_{1}A_{2}B_{1}-\frac{\tilde{h} _{23}}{4}A_{2}A_{3}B_{3}\right.  \nonumber \\
& \qquad \left.
-\frac{\tilde{h}_{12}}{8}B_{2}(A_{1}^{2}+3B_{1}^{2})-\frac{
\tilde{h}_{23}}{8}B_{2}(A_{3}^{2}+3B_{3}^{2})\right] +O(\epsilon
^{2})\,,
\nonumber \\
A_{3}^{\prime }& =\frac{\epsilon }{\kappa }\left[ \left(
\frac{\tilde{\mu} _{1}}{2}-\frac{\tilde{V}_{0}}{4}\right)
B_{3}-\frac{3\tilde{g}}{8}
B_{3}(A_{3}^{2}+B_{3}^{2})-\frac{\tilde{\alpha}_{13}}{2}B_{1}-\frac{\tilde{
\alpha}_{23}}{2}B_{2}-\frac{\tilde{h}_{13}}{4}A_{1}A_{3}B_{1}-\frac{\tilde{h} _{23}}{4}A_{2}A_{3}B_{2}\right.  \nonumber \\
& \qquad \left.
-\frac{\tilde{h}_{13}}{8}B_{3}(A_{1}^{2}+3B_{1}^{2})-\frac{
\tilde{h}_{23}}{8}B_{3}(A_{2}^{2}+3B_{2}^{2})\right] +O(\epsilon
^{2})\,,
\nonumber \\
B_{1}^{\prime }& =\frac{\epsilon }{\kappa }\left[ -\left(
\frac{\tilde{\mu }_{1}}{2}+\frac{\tilde{V}_{0}}{4}\right)
A_{1}+\frac{3\tilde{g}}{8}
A_{1}(A_{1}^{2}+B_{1}^{2})+\frac{\tilde{\alpha}_{12}}{2}A_{2}+\frac{\tilde{
\alpha}_{13}}{2}A_{3}+\frac{\tilde{h}_{12}}{4}A_{2}B_{1}B_{2}+\frac{\tilde{h} _{13}}{4}A_{3}B_{1}B_{3}\right.  \nonumber \\
& \qquad \left.
+\frac{\tilde{h}_{12}}{8}A_{1}(3A_{2}^{2}+B_{2}^{2})+\frac{
\tilde{h}_{13}}{8}A_{1}(3A_{3}^{2}+B_{3}^{2})\right] +O(\epsilon
^{2})\,,
\nonumber \\
B_{2}^{\prime }& =\frac{\epsilon }{\kappa }\left[ -\left(
\frac{\tilde{\mu }_{1}}{2}+\frac{\tilde{V}_{0}}{4}\right)
A_{2}+\frac{3\tilde{g}}{8}
A_{2}(A_{2}^{2}+B_{2}^{2})+\frac{\tilde{\alpha}_{12}}{2}A_{1}+\frac{\tilde{
\alpha}_{23}}{2}A_{3}+\frac{\tilde{h}_{12}}{4}A_{1}B_{1}B_{2}+\frac{\tilde{h} _{23}}{4}A_{3}B_{2}B_{3}\right.  \nonumber \\
& \qquad \left.
+\frac{\tilde{h}_{12}}{8}A_{2}(3A_{1}^{2}+B_{1}^{2})+\frac{
\tilde{h}_{23}}{8}A_{2}(3A_{3}^{2}+B_{3}^{2})\right] +O(\epsilon
^{2})\,,
\nonumber \\
B_{3}^{\prime }& =\frac{\epsilon }{\kappa }\left[ -\left(
\frac{\tilde{\mu }_{1}}{2}+\frac{\tilde{V}_{0}}{4}\right)
A_{3}+\frac{3\tilde{g}}{8}
A_{3}(A_{3}^{2}+B_{3}^{2})+\frac{\tilde{\alpha}_{13}}{2}A_{1}+\frac{\tilde{
\alpha}_{23}}{2}A_{2}+\frac{\tilde{h}_{13}}{4}A_{1}B_{1}B_{3}+\frac{\tilde{h} _{23}}{4}A_{2}B_{2}B_{3}\right.  \nonumber \\
& \qquad \left.
+\frac{\tilde{h}_{13}}{8}A_{3}(3A_{1}^{2}+B_{1}^{2})+\frac{
\tilde{h}_{23}}{8}A_{3}(3A_{2}^{2}+B_{2}^{2})\right] +O(\epsilon
^{2})\,. \label{res3}
\end{align}

One can find double-mode solutions to (\ref{res3}) that are
analogous to those of Eq. (\ref{res}). For example, if $|\alpha
_{13}|=|\alpha _{23}|$, so that the first and second components in
the ternary condensate have the same strength in their linear
coupling to the third component, there exists a double-mode
equilibrium with $A_{1}^{2}=A_{2}^{2}$ and $
A_{3}=B_{1}=B_{2}=B_{3}=0$. The values of $A_{1}$ and $A_{2}$ are
exactly as for binary BECs [see Eq. (\ref{a1a2type1})], except
that $\alpha $ and $h $ in the solution are replaced by $\alpha
_{12}$ and $h_{12}$. Further, in this case, one finds
$A_{1}=-A_{2}$ for $\alpha _{13}=\alpha _{23}$ and $ A_{1}=A_{2}$
for $\alpha _{13}=-\alpha _{23}$. In fact, these modes are a
straightforward extension of their two-component counterparts, as
the third component is absent in the stationary solution.
Furthermore, the stability of the symmetric double-mode
equilibria, reported above, ensures the stability of these
solutions in the ternary model.

\begin{figure}[tbp]
{\epsfig{file=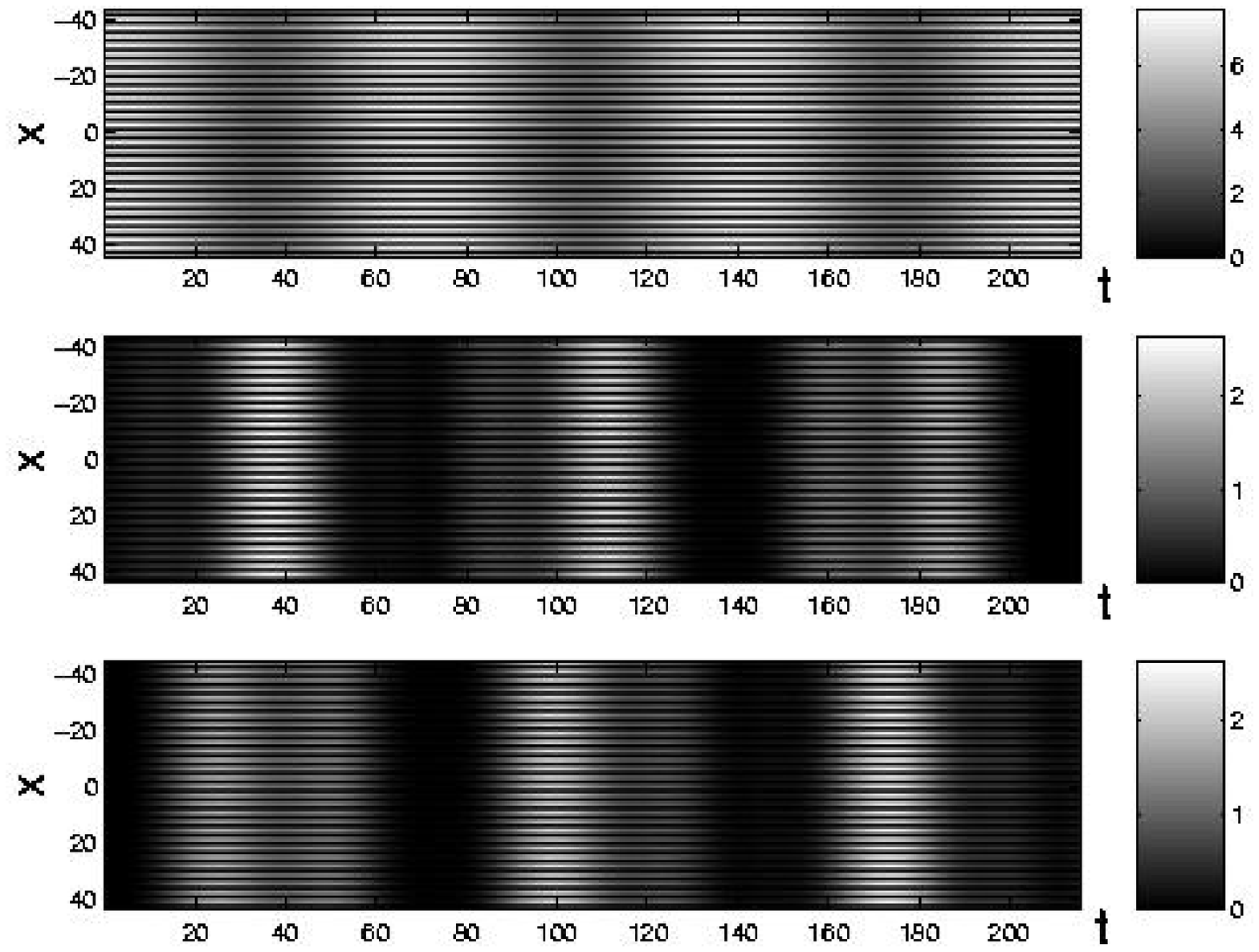, width=7.5cm,angle=0, clip=}}
{\epsfig{file=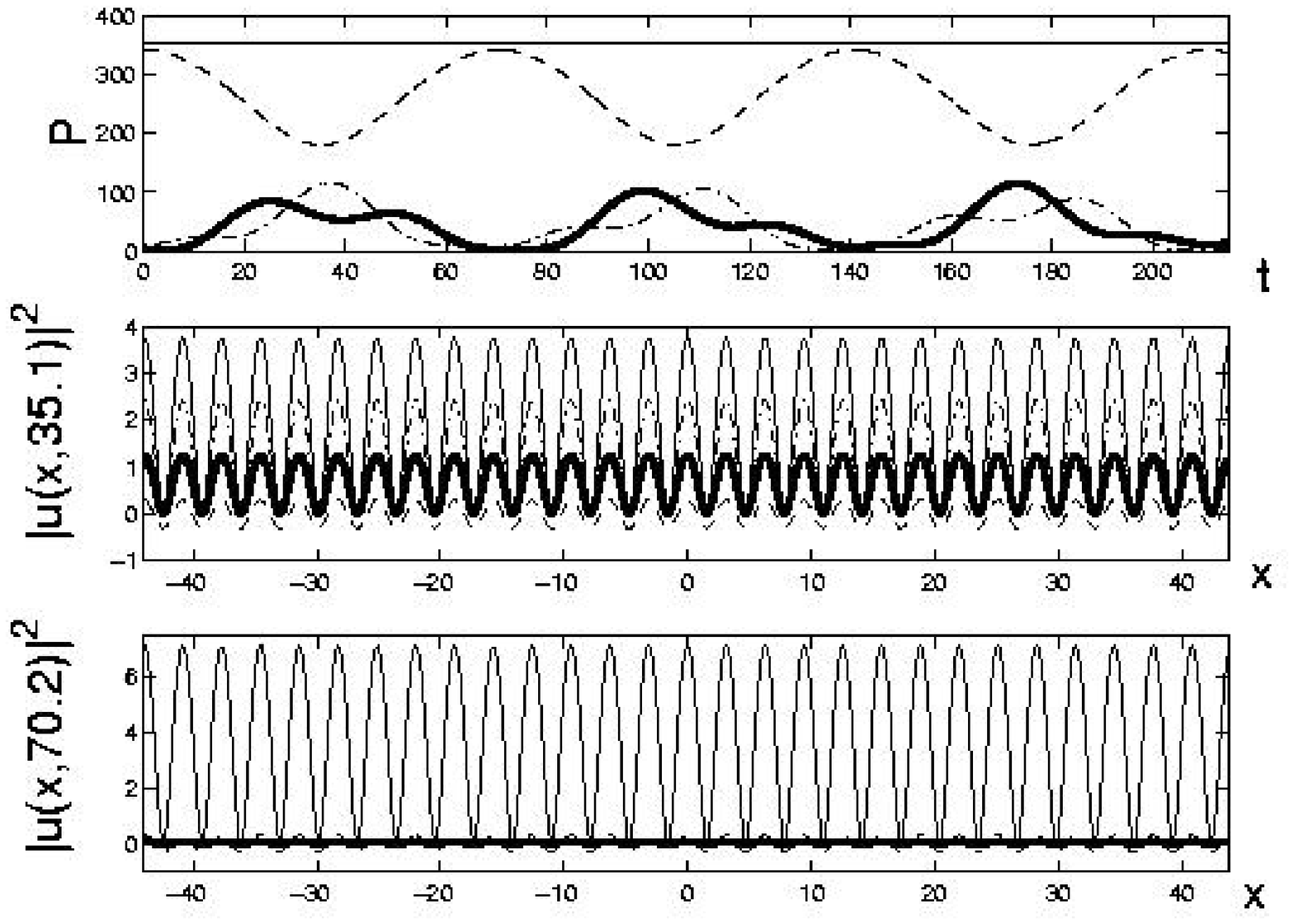,width=7.5cm,angle=0, clip=}}
\caption{The same as in Fig. \protect\ref{mpfig4}, but for the
three-component (ternary) model. The bottom panel of the left plot
shows the spatio-temporal evolution of the third component (which
is absent in the unperturbed, unstable two-mode solution), and
the thick solid line on the right shows the evolution of the
number of atoms in this component (top subplot). Its spatial
profile is shown as well (middle and bottom subplots, for $ t=35.1$
and $t=70.2$, respectively). Periodic oscillations of matter
between {\em all three components} are evident, cf. the
oscillations in the two-compoment model, displayed in Fig.
\protect\ref{mpfig4}. The parameters are $g=0.025$, $h=0.005$,
$\protect\alpha _{12}=\protect\alpha _{13}=-0.02$, $V_{0}=0.3$,
and $\protect\kappa =\protect\mu =1$. The quantity $\protect\alpha
_{23}\approx 0.117$ is determined from the values of $\protect\alpha
_{12}=\protect\alpha _{13}=-0.02$ (see text).} \label{mpfig6}
\end{figure}

The situation is more interesting for {\em asymmetric} two-mode
solutions, such as the ones corresponding to Eq. (\ref{unequal}),
which are, simultaneously, solutions to Eqs. (\ref{cnls3}) with
$A_{3}=B_{3}=0$, provided $ A_{1}=-(\alpha _{23}/\alpha
_{13})A_{2}$.  [Note that this relation is used to determine
$\alpha_{23}/\alpha_{13}$, as $A_1$ and $A_2$ are determined from
Eq. (\ref{unequal}).]  Direct simulations of the three-component GP
equations (\ref{cnls3}) with $h := h_{12} = h_{13} = h_{23}$ show that these
asymmetric solutions are unstable,
just as in the two-component model. The instability
development, illustrated by Fig. \ref{mpfig6}, leads to an
interesting dynamical interplay between the components. In
particular, as a result of the instability, the third component is
eventually excited, which leads to periodic oscillation of matter
between all three components; i.e., in this case, we observe a
true example of three-component dynamics.

\section{Conclusions}

In this work, we analyzed spatial structures in coupled Gross-Pitaevskii
(coupled GP) equations, which include both nonlinear and linear
interactions, in an optical-lattice (OL) potential. The model describes a
BEC consisting of a mixture of two different hyperfine states of one atomic
species, which are linearly coupled by a resonant electromagnetic field. In
the absence of the OL, we found plane-wave solutions and examined their
stability. In the presence of the OL, we derived a system of averaged
equations to describe a spatially modulated state which is coupled to the
periodic potential through a subharmonic resonance. We found equilibria of 
the latter system and examined the stability of the corresponding spatially 
periodic solutions to the coupled GP equations using direct simulations.
We demonstrated that symmetric dual-mode resonant states with two equal
amplitudes are stable, whereas asymmetric ones (with unequal amplitudes) are
unstable, generating solutions that oscillate periodically in time. The latter
type of dynamical behavior is only possible in the presence of linear
coupling between BEC components.  We found four-mode stationary solutions as 
well, but they are always unstable. Finally, a three-component generalization
of the model was introduced and briefly considered. In this case, we found
that the unstable asymmetric two-mode solution, with one component originally
empty, develops time-periodic oscillations in which the initially empty
component becomes populated.

\section*{Acknowledgements}

We appreciate useful discussions with Bernard Deconinck, Alex Kuzmich,
Alexandru Nicolin, and Richard Rand. P.G.K. gratefully acknowledges support
from NSF-DMS-0204585, from the Eppley Foundation for Research and from an
NSF-CAREER award. The work of B.A.M. was supported in a part by the grant
No. 8006/03 from the Israel Science Foundation.

\section*{Appendix}

The functions $F_{A_{j}}$ and $F_{B_{j}}$, which appear in Eqs.
(\ref{preavg}), can be written as a sum of harmonic
contributions. To simplify the notation, we write $\tilde{g}$,
$\tilde{h}$, $\tilde{\alpha}$, and $\tilde{V} _{0}$ simply as $g$,
$h$, $\alpha $, and $V$.

In the non-resonant case, $F_{A_{1}}=G_{A_{1}}/\sqrt{\mu }$, where
\begin{align}
G_{A_{1}}(A_{1},A_{2},A_{3},A_{4},x)& =\left[ -\frac{\alpha }{2}-\frac{3g}{8} B_{1}(A_{1}^{2}+B_{1}^{2})-\frac{h}{4}A_{1}A_{2}B_{2}-\frac{h}{8} B_{1}(A_{2}^{2}+3B_{2}^{2})\right]   \nonumber \\
& \quad +\left[ -\frac{\alpha
}{2}-\frac{g}{4}A_{1}(A_{1}^{2}+3B_{1}^{2})-
\frac{h}{2}A_{2}B_{1}B_{2}-\frac{h}{4}A_{1}(A_{2}^{2}+B_{2}^{2})\right]
\sin
(2\sqrt{\mu }x)  \nonumber \\
& \quad +\left[ \frac{g}{8}A_{1}(3B_{1}^{2}-A_{1}^{2})+\frac{h}{4}
A_{1}B_{1}B_{2}+\frac{h}{8}A_{1}(B_{2}^{2}-A_{2}^{2})\right] \sin
(4\sqrt{
\mu }x)  \nonumber \\
& \quad +\left[ -\frac{V}{4}A_{1}\right] \sin (2[\kappa
-\sqrt{\mu }]x)+
\left[ \frac{V}{4}\right] \sin (2[\kappa +\sqrt{\mu }]x)  \nonumber \\
& \quad +\left[ \frac{g}{2}B_{1}^{3}+\frac{\alpha }{2}B_{2}+\frac{h}{2} B_{1}B_{2}^{2}\right] \cos (2\sqrt{\mu }x)  \nonumber \\
& \quad +\left[
\frac{g}{8}B_{1}(3A_{1}^{2}-B_{1}^{2})+\frac{h}{4}A_{1}A_{2}B_{2}+
\frac{h}{8}B_{1}(A_{2}^{2}-B_{2}^{2})\right] \cos (4\sqrt{\mu
}x)
\nonumber \\
& \quad +\left[ \frac{V}{2}B_{1}\right] \cos (2\kappa x)+\left[
-\frac{V}{4} \right] \cos (2[\kappa -\sqrt{\mu }]x)+\left[
-\frac{V}{4}\right] \cos (2[\kappa +\sqrt{\mu }]x)\,,
\label{harm1}
\end{align}
and $F_{B_{1}}=G_{B_{1}}/\sqrt{\mu }$, where
\begin{align}
G_{B_{1}}(A_{1},A_{2},A_{3},A_{4},x)& =\left[ \frac{\alpha
}{2}A_{2}+\frac{3g
}{8}A_{1}(A_{1}^{2}+B_{1}^{2})+\frac{h}{4}A_{2}B_{1}B_{2}+\frac{h}{8} A_{1}(A_{2}^{2}+B_{2}^{2})\right]   \nonumber \\
& \quad +\left[ \frac{\alpha }{2}B_{2}+\frac{g}{4} B_{1}(3A_{1}^{2}+B_{1}^{2})+\frac{h}{2}A_{1}A_{2}B_{2}+\frac{h}{4} B_{1}(A_{2}^{2}+B_{2}^{2})\right] \sin (2\sqrt{\mu }x)  \nonumber \\
& \quad +\left[ \frac{g}{8}B_{1}(3A_{1}^{2}-B_{1}^{2})+\frac{h}{4}
A_{1}A_{2}B_{2}+\frac{h}{8}B_{1}(A_{2}^{2}-B_{2}^{2})\right] \sin
(4\sqrt{
\mu }x)  \nonumber \\
& \quad +\left[ \frac{V}{4}B_{1}\right] \sin (2[\kappa
-\sqrt{\mu }]x)+
\left[ -\frac{V}{4}\right] \sin (2[\kappa +\sqrt{\mu }]x)  \nonumber \\
& \quad +\left[ \frac{\alpha }{2}A_{2}+\frac{g}{2}A_{1}^{3}+\frac{h}{2} A_{1}A_{2}^{2}\right] \cos (2\sqrt{\mu }x)  \nonumber \\
& \quad +\left[
\frac{g}{8}A_{1}(A_{1}^{2}-3B_{1}^{2})-\frac{h}{4}A_{2}B_{1}B_{2}+
\frac{h}{8}A_{1}(A_{2}^{2}-B_{2}^{2})\right] \cos (4\sqrt{\mu
}x)
\nonumber \\
& \quad +\left[ -\frac{V}{2}A_{1}\right] \cos (2\kappa x)+\left[
-\frac{V}{4} A_{1}\right] \cos (2[\kappa -\sqrt{\mu }]x)+\left[
-\frac{V}{4}A_{1} \right] \cos (2[\kappa +\sqrt{\mu }]x)\,.
\label{harm2}
\end{align}
Only $O(1)$ [i.e., constant harmonic] terms remain after averaging.

In the resonant case, one obtains, after averaging, an extra term
depending on the periodic potential $V$, because a term that was a
prefactor of a non-constant harmonic in (\ref{harm1}) and
(\ref{harm2}) has become a coefficient in front of the $O(1)$
term. Other harmonic terms are also simplified due to the
resonance, but they nevertheless do not contribute to the averaged
equations because they are still prefactors of
non-constant harmonics. The extra terms with $\mu _{1}$ arise
from Taylor expanding in powers of $\epsilon $ and keeping the
leading-order terms. In the resonant case,
$F_{A_{1}}=G_{A_{1}}/\kappa $ and $F_{B_{1}}=G_{B_{1}}/ \kappa $.

In both the resonant and non-resonant cases, the expressions for $F_{A_2}$
and $F_{B_2}$ are obtained by switching the subscripts $1
\longleftrightarrow 2$ in the equations above.


\end{document}